\documentclass[a4paper,10pt,journal]{IEEEtran}
\usepackage{cite}
\usepackage[binary-units=true]{siunitx}
\DeclareSIUnit \GHz {GHz}
\usepackage[pdftex]{graphicx}
\DeclareSIUnit \dBm {dBm}
\DeclareSIUnit \dB {dB} 
\DeclareSIUnit \dBi {dBi} 
\DeclareSIUnit \Kbps {Kbps}
\DeclareSIUnit \Mbps {Mbps}
\DeclareSIUnit \Gbps {Gbps}
\DeclareSIUnit \kBps {kBps}
\DeclareSIUnit \MBps {MBps}
\DeclareSIUnit \GBps {GBps}

\usepackage{amsfonts}
\usepackage{amsmath}
\usepackage[caption=false,font=footnotesize]{subfig}
\usepackage[hyphens]{url}
\usepackage{algorithm}
\usepackage{algpseudocode}

\newcommand{\FF}[1]{{\mathbb{F}}}

\usepackage{tikz}
\usetikzlibrary{arrows,automata}

\usepackage{enumerate}

\newcolumntype{P}[1]{>{\centering\arraybackslash}p{#1}}
\newcolumntype{M}[1]{>{\centering\arraybackslash}m{#1}}
\begin{document}
\title{Agile Data Offloading over Novel\\ Fog Computing Infrastructure for CAVs}
\author{\IEEEauthorblockN{Andrea Tassi\IEEEauthorrefmark{1}, Ioannis Mavromatis\IEEEauthorrefmark{1}, Robert Piechocki\IEEEauthorrefmark{1}\IEEEauthorrefmark{2}, Andrew Nix\IEEEauthorrefmark{1},\\ Christian Compton\IEEEauthorrefmark{3}, Tracey Poole\IEEEauthorrefmark{3} and Wolfgang Schuster\IEEEauthorrefmark{3}}\\ 
\IEEEauthorblockA{\IEEEauthorrefmark{1}Department of Electric and Electronic Engineering, University of Bristol, UK}\\
\IEEEauthorblockA{\IEEEauthorrefmark{2}The Alan Turing Institute, London, NW1 2DB, UK}\\
\IEEEauthorblockA{\IEEEauthorrefmark{3}Atkins Global Limited, London, UK}}

\maketitle

\begin{abstract}
Future Connected and Automated Vehicles (CAVs) will be supervised by cloud-based systems overseeing the overall security and orchestrating traffic flows. Such systems rely on data collected from CAVs across the whole city operational area. This paper develops a Fog Computing-based infrastructure for future Intelligent Transportation Systems (ITSs) enabling an agile and reliable off-load of CAV data. Since CAVs are expected to generate large quantities of data, it is not feasible to assume data off-loading to be completed while a CAV is in the proximity of a single Road-Side Unit (RSU). CAVs are expected to be in the range of an RSU only for a limited amount of time, necessitating data reconciliation across different RSUs, if traditional approaches to data off-load were to be used. To this end, this paper proposes an agile Fog Computing infrastructure, which interconnects all the RSUs so that the data reconciliation is solved efficiently as a by-product of deploying the Random Linear Network Coding (RLNC) technique. Our numerical results confirm the feasibility of our solution and show its effectiveness when operated in a large-scale urban testbed.
\end{abstract}

\begin{IEEEkeywords}Fog Computing, ITS, CAV, V2X, Network Coding,  Data Offloading.\end{IEEEkeywords}

\section{Introduction}
Connected and Autonomous Vehicles (CAVs) will play a pivotal role in our everyday lives in the future. CAVs, fully integrated with sensors, will be able to provide awareness of the surrounding environment and coordination with fixed network nodes and nearby vehicles~\cite{qosRequirements}. To do so, they will require to exchange a vast amount of data, accommodated over various means of communication frameworks, and operated in a heterogeneous fashion~\cite{hetNet,TassimmWave}. The broadcast and public nature of these communication links make them susceptible to cyber-security threats, and hence, impacting the passengers' safety. The vulnerability of automotive systems was recently demonstrated by the Chrysler Jeep hack, where the attackers exploited breaches in the vehicles internet-connected entertainment system\footnote{https://www.bbc.co.uk/news/technology-33650491}.

Responding effectively to cyber-security incidents requires a number of different technological, procedural and organizational elements to be in place and work effectively. In many ways, the cyber-security requirements of CAVs are like any other business critical or safety-related system. Commercial Off-The-Shelf (COTS) equipment usually provides its information in forms of events and alarms. Correlation can be used to turn this data into more meaningful information. Machine Learning and Artificial Intelligence could be used to detect anomalies, which could affect the system~\cite{machineLearning}. New devices (such as roadside devices) could pass information (data and security oriented) to a centrally managed system~\cite{fogArchitecture,fogComputing}. The authenticity and integrity of that information is of paramount importance.
%What is more, understanding what can be trusted, is a principal action. 
Being empowered by this knowledge and having the capability of detecting when the system is compromised, allow system administrators to determine the right course of actions~\cite{anomalyDetectionActions}.

In this paper, we focus on the pressing concern of empowering cloud-based services to detect incidents and respond to them by defining an agile strategy for data collection. In particular, we will propose a novel data offloading strategy allowing CAVs to share their sensory information with the Fog Computing Infrastructure and then, with cloud-based services. We envisage a system where our network will incorporate Fog Computing capabilities, giving us the leverage to use powerful processing nodes one-hope away from each Road-Side Unit (RSU). This will significantly reduce the latency of the data processing, minimizing the time required to identify potential threats. 

The data offloading from a CAV to an RSU takes place when CAVs are in within range with one or more RSUs. The sensor data exchange happens in a broadcast fashion with the receivers and is organized as streams of information. However, the heterogeneity in driving behaviors within a city and the sparsity of the RSUs lead to intermittent communications. Reconciling the data at the infrastructure level becomes highly inefficient. Employing network coding techniques for achieving secure systems has attracted the interest of the research community lately, especially for wireless broadcast applications~\cite{networkCoding} and secure vehicular networks~\cite{networkCodingVehicular}. Inspired by that, and taking into account the fragmentary nature of the data offloading, we propose a new framework that uses Random Linear Network Coding (RLNC) to achieve an agile data offloading in Intelligent Transportation Systems (ITSs).

More specifically, our idea is based on the existing ITS-G5 Dedicated Short Range Communication (DSRC) protocol stack~\cite{etsiStandard}. We propose an amendment as well as a RLNC version of the Cooperative Awareness Message (CAM), that is expected to be used for broadcasting sensor data. The amendment takes place at the Facilities layer. Later in this work, we will describe the proposed sublayer and its functionality. The coded packets generated from this sublayer are then mapped onto the proposed RLNC-CAM and broadcast to the network. When received from a RSU, the Fog Computing Infrastructure and the RLNC decoder are responsible for decoding the packet that could be then be forwarded to other cloud-based services. 

The rest of the paper is organized as follows. Sec.~\ref{sec:SM} describes our system model. The proposed Fog Computing implementation and Cloud Computing capabilities are presented at first. Then, the need for RLNC and how it could operate within the context of a vehicular network are introduced. Following, we present the proposed extension to the ITS-G5 stack (Sec.~\ref{sec:proposedExt}). More specifically, we start by describing the new RLNC-CAM message and the information encapsulated in that. In addition, we outline the design of the RLNC-Facility sublayer responsible for the generation of RLNC-CAMs. In Sec.~\ref{sec:perfEval}, we present our numerical results that confirm the feasibility of our proposal. We focused our performance investigation on real-world data traces captured during an extensive experimental campaign that took place in the City of Bristol, UK. Finally, we summarize our findings in Sec.~\ref{sec:conclusions} and present a critical analysis of our results.

\section{System Model}\label{sec:SM}
We consider a city-scale network providing wireless connectivity to CAVs. In our system model, each CAV wishes to offload its sensing acquired data onto a cloud-store facility by means of RSUs connected to a network of Fog computing infrastructure nodes, called Fog Orchestrators (FOs).

\subsection{Fog Computing Infrastructure for Cyber-Secure ITSs}
We assume our network being clustered in different management areas called Fog Areas. Each Fog Area consists of a number of RSUs, all being connected to a particular FO. We also assume that one FO serves each Fog Area. The FOs represents the logical entities encapsulating the core components of our system.
%The number of constraints in an FO and its building blocks (e.g., microservices) can be many.
%The FO is responsible for orchestrating them in either a centralized or distributed manner while maintaining its core functionality. 
FO ensures the programmability of the system simplifying the deployment of new functionalities, which rely upon a network of sensors and actuators. 

In this particular system model, the FO is responsible for collecting the transmitted RLNC-CAMs, removing duplicated messages that may arise as a CAV is in range with two or more RSUs, recovering the encapsulated message, and passing them to cloud services. There, the threats can be effectively detected, and cyber-security decisions are taken. %All the above ensure the QoS required for a specific cyber-security service.

\begin{figure*}[t]     
\centering
\includegraphics[width=0.8\textwidth]{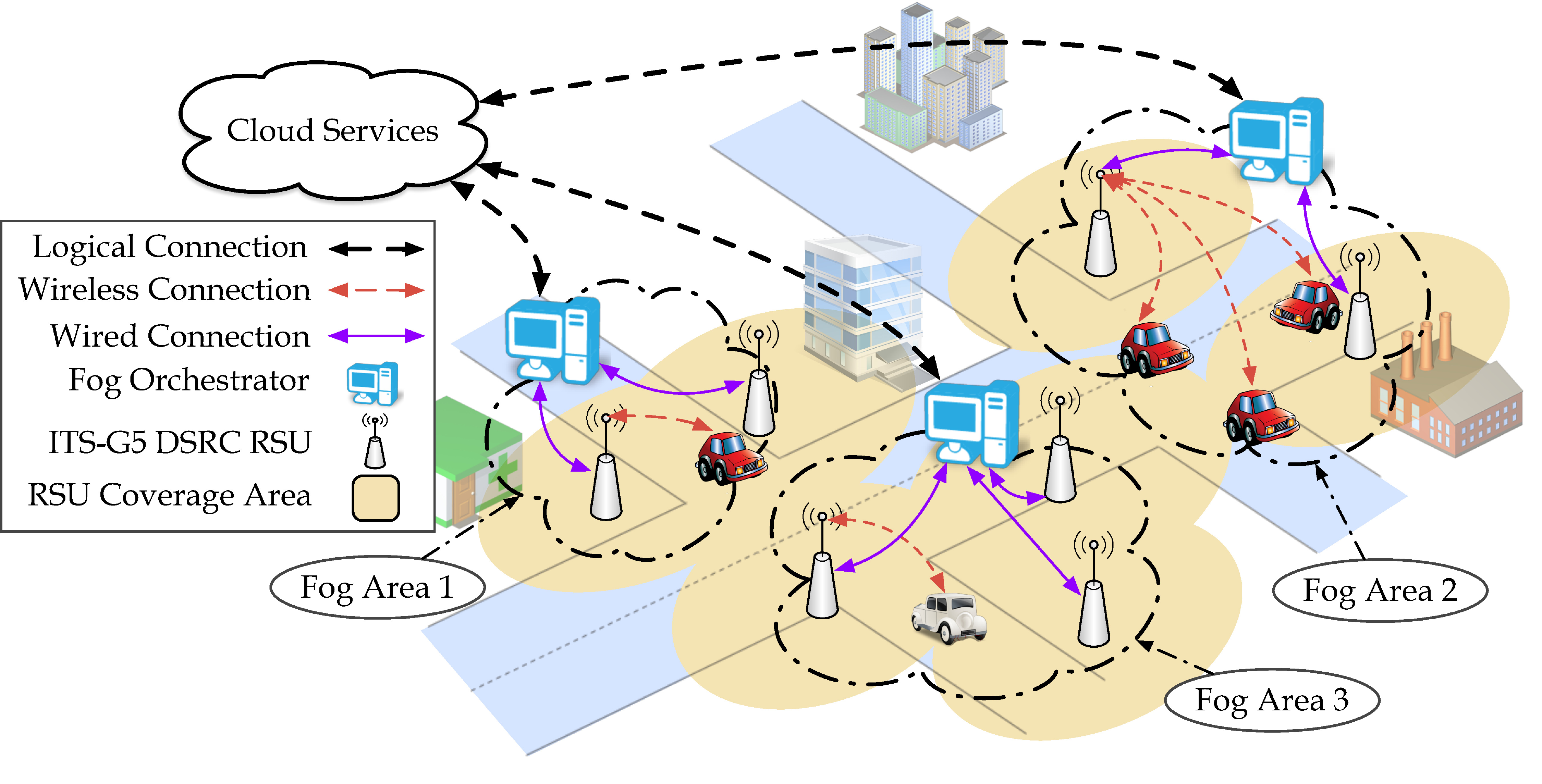}
    \caption{General overview of the considered Fog Computing infrastructure for CAVs.}
    \label{fig:fogNW}
\end{figure*}

We assume that a cloud-based data controller is responsible for the processing of all the sensor data transmitted by each CAV. Ultimately, RSUs are solely responsible for relaying messages to and from the CAVs. As shown in Fig.~\ref{fig:fogNW}, this solution interacts with a cloud-based city-wide connection. In particular, the cloud-based service will only be in charge of recording city-scale data, interconnecting the different Fog Areas and enforcing city-scale policies to be put into practice.

This system model provides the necessary abstraction for a next-generation ITS system. This will also give us the leverage to deliver cyber-secure ITSs. In fact, as CAV technologies come to play an ever-increasing role in the safe and smooth running of road networks, network operators will increasingly need to implement effective cyber-security measures. These measures need to ensure that road users’ personal data is not compromised but just as importantly (arguably more importantly) they must ensure that the systems which CAVs rely on work as designed. An attacker could take more focused actions, which could result in an even greater impact on journeys, they may choose to spoof communications to systems for example which could then possibly have an impact on safety. 

\subsection{RLNC for Agile Data Offloading}\label{sec:SM.RLNC}
For the data offloading to happen, each CAV broadcasts its sensor data as it gets within range with one or more RSUs.
%Sensor data is organized as a stream of information packets.
However, as the CAV leaves the coverage area of a RSU, the data offloading process may be interrupted to be resumed as the CAV reaches the next coverage region. This particularly applies, in sparsely deployed RSU networks. 
%However, as the CAV enters and leaves the coverage area of an RSU, the data offloading process may be interrupted for several seconds or even hours before the CAV being in range with another RSU. In particular, this applies when RSUs are sparsely deployed. 
For this reason, traditional handover procedures cannot be put in place -- thus making the reconciliation of data received by multiple RSUs very inefficient and not scalable, as the number of CAVs increases. With these regards, RLNC promises to overcome these issues in a seamless fashion~\cite{6416071}.

Consider the case of a CAV wishing to transmit to one or multiple RSUs a source message over a broadcast channel. In our system, a source message represents a stream of sensor information packets. Transmissions experience a certain packet error probability. According to the RLNC principle, the CAV segments the source message into $K$ source packets $\{\mathbf{s}_i\}_{i=1}^K$ where $\mathbf{s}_i$ is made of $L$ elements from a finite field $\mathbb{F}_q$. The CAV also linearly combines at random the source packets to obtain $N$ coded packets $\mathcal{C}=\{\mathbf{c}_i\}_{i=1}^{N}$ for transmission. Each coded packet $\mathbf{c}_j$ consists of $L$ elements from $\mathbb{F}_q$ and it is obtained as:
\begin{equation}
\mathbf{c}_j = \sum_{i = 1}^K g_{i,j} \cdot \mathbf{s}_i,
\end{equation}
where the coding coefficient $g_{i,j} \in \mathbb{F}_q$. Besides, let $\mathbf{G}$ be a $K \times N$ random matrix over $\mathbb{F}_q$ where its $(i,j)$-th element is defined by $g_{i,j}$, the relation
\begin{equation}
[\mathbf{c}_1, \mathbf{c}_2, \ldots, \mathbf{c}_{N}] = [\mathbf{s}_1, \mathbf{s}_2, \ldots, \mathbf{c}_K] \cdot \mathbf{G}    
\end{equation}
holds true.

Let $\mathcal{C}$ be the set of $n$ coded packets pertaining to the same CAV and successfully received by each of the RSU, for $0 \geq n \geq N$. In turn, each RSU forwards to the FO and then to a cloud-based service the received coded packets. Hence, the cloud-based service can populate a $K \times n$ decoding matrix where the $n$ columns of $\mathbf{M}$ are defined by the $n$ columns in $\mathbf{G}$ associated with coded packets in $\mathcal{C}$. The source message is recovered as soon as the rank of $\mathbf{M}$ becomes equal to $K$. In particular, the probability $\mathrm{R}(n)$ of a source message of being recovered, as a function of $n$, can be expressed follows~\cite{7335581}:
\begin{equation}\label{eq:recoveryProb}
    \mathrm{R}(n) = \prod_{t = 0}^{K-1} \left( 1 - \frac{1}{q^{n-t}} \right)
\end{equation}

In our system, the transmission of each source message takes place in an unacknowledged fashion. After that the $N$-th coded packet associated with a source message has been transmitted by a CAV, the broadcasting of the following source packet begins. In the following section, we will describe how a sensor data stream can be partitioned into a sequence of source message and how the RLNC principle can be integrated into the ETSI's ITS-G5 communication stack.

\section{Proposed Extension to the ITS-G5 Stack}\label{sec:proposedExt}
\begin{figure}[t]     
\centering
\includegraphics[width=0.8\columnwidth]{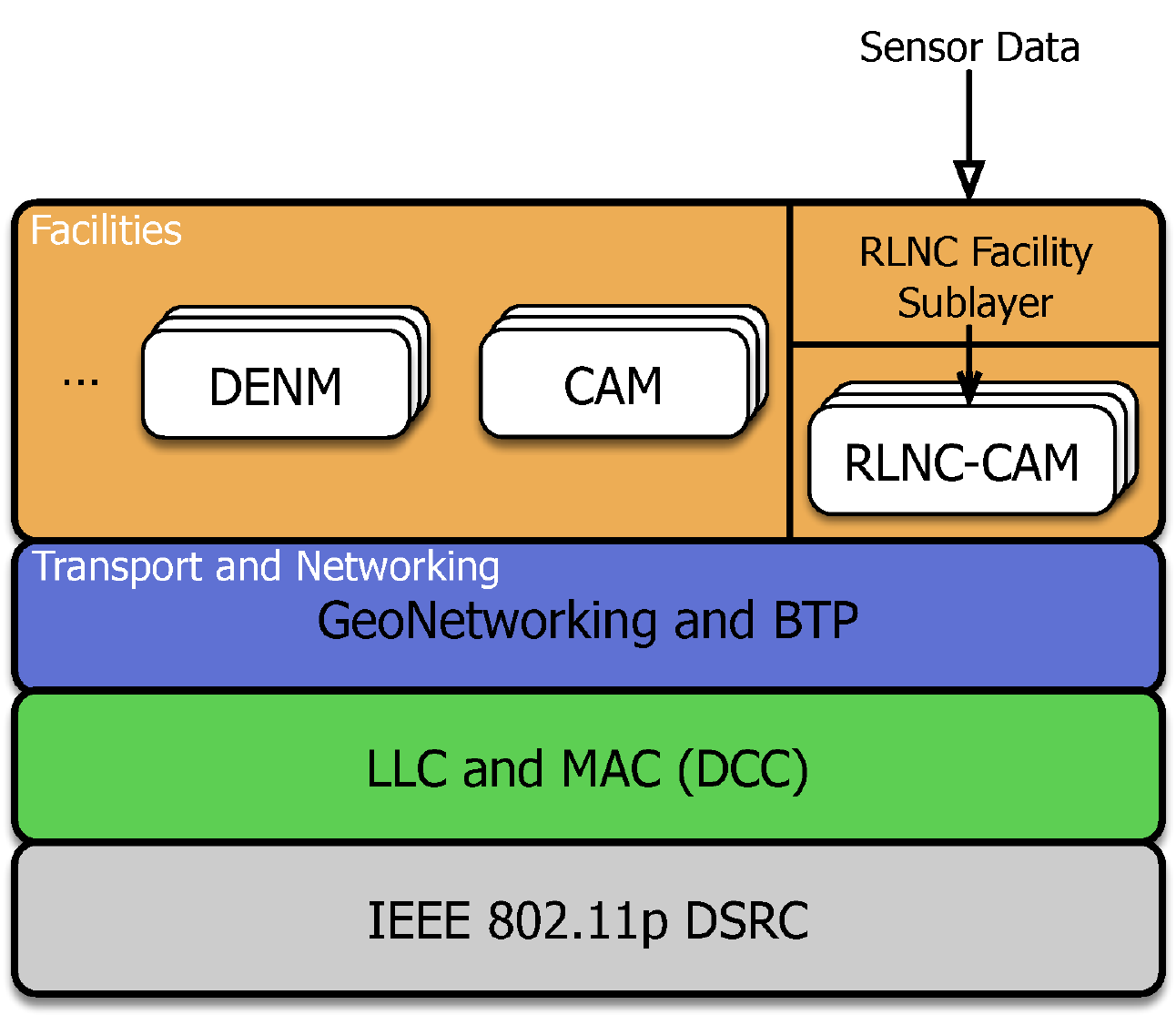}
    \caption{Proposed amendment to the ETSI's ITS-G5 protocol stack including the RLNC-Facility sublayer.}
    \label{fig:stack}
\end{figure}

In our system, we assume that CAVs and RSUs communicate using the ETSI's ITS-G5 standard~\cite{etsiStandard}. The ITS-G5 standard has been derived from the US Wireless Access in Vehicular Environment (WAVE), and both of them build upon the IEEE 802.11p DSRC physical layer. As shown in Fig.~\ref{fig:stack}, the Medium Access Control (MAC) layer of the ITS-G5 protocol stack employs the Decentralized Congestion Control (DCC) protocol to dynamically optimize the channel load. This is done by adapting the transmission power, the rate and the sensitivity of a transceiver. Then a simplified Logical Link Control (LLC) layer acts as an adaptation layer between the DCC and the upper layers. With regards to the Networking and Transport Layer, packets can be forwarded in a multi-hop fashion by the GeoNetworking protocol, which employs the Contention Based Forwarding (CBF) algorithm. In particular, on the basis of geolocation information, the CBF algorithm elects as multi-point relay node the CAV at the greatest distance from the source node. On the other hand, point-to-point communications are handled by the Basic Transport Protocol (BTP), which is responsible for multiplexing/demultiplexing packets originated by the Facility layer~\cite{6979970}.

At the level of the Facility layer, CAMs and Decentralized Environmental Messages (DENMs) are transmitted/received. In particular, CAMs are periodically broadcast by each CAV and convey information pertaining to the position of a CAV, its engine status, etc. On the other hand, DENMs are transmitted in the response of an event and are used to alert road users for a danger ahead, adverse weather conditions, etc~\cite{6979970}. As per Sec.~\ref{sec:SM}, in order to enable cloud-based services to detect cyber-security threats efficiently, CAVs are expected to share their sensor information periodically. As such, building upon the standard definition of a CAM, we propose the RLNC-CAM version that each CAV is expected to use to broadcast sensor information. In particular, as shown in Fig.~\ref{fig:stack}, when a CAV wishes to employ RLNC-CAMs, sensor information is provided as an input to the RLNC-Facility sublayer, which is responsible for the following:
\begin{itemize}
    \item Segmenting the sensor data stream into a sequence of source packets with the same bit length.
    \item Organizing the sequence of source packets into source messages defined by $K$ consecutive source packets.
    \item Assigning an ID to each source message.
    \item Generating a sequence of $N$ coded packets per each source message according to the RLNC principle (see Sec.~\ref{sec:SM.RLNC}).
\end{itemize}

\begin{figure}[t]     
\centering
\includegraphics[width=0.8\columnwidth]{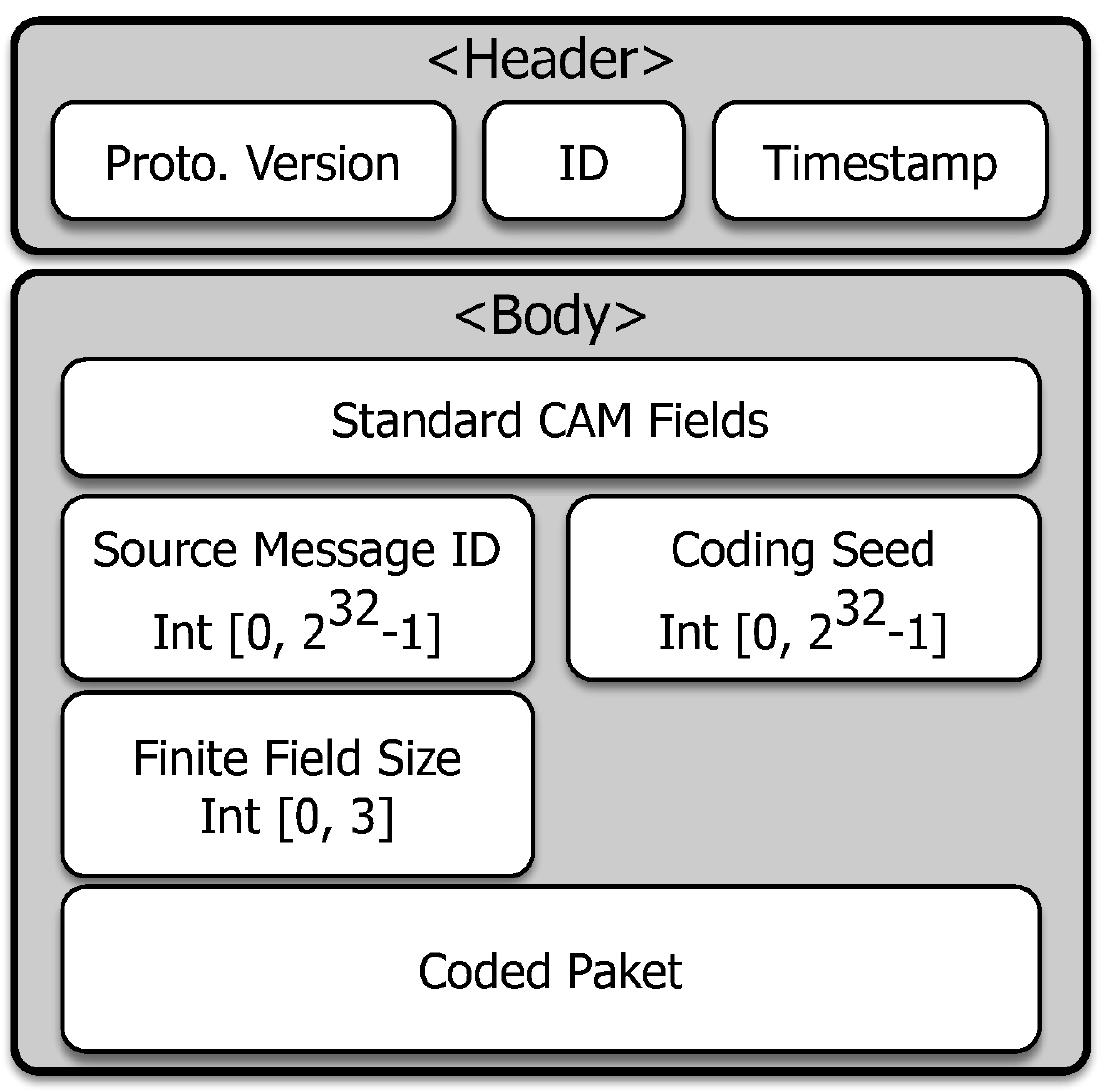}
    \caption{Structure of the proposed RLNC-CAM message. Both the standard CAM fields and the RLNC fields are encapsulated in the body of the message.}
    \label{fig:cam}
\end{figure}

Each coded packet generated by the RLNC-Facility Sublayer is mapped into an RLNC-CAM having the structure shown in Fig.~\ref{fig:cam}. Each RLNC-CAM comprises the same fields as a standard CAM~\cite{etsiCam}. In addition, the RLNC fields are added to the message. More specifically, the header of an RLNC-CAM includes the protocol version in use, a general CAM ID and a generation timestamp. The body hosts information pertaining the CAV transmitting the message, such as: a transmitter ID, the nature of the CAV (mobile, public authority, private, etc.), its position (latitude, longitude, elevation, and heading) and a set of optional attributes.

\begin{figure*}[ht]
\minipage{0.49\textwidth}
\centering
    \includegraphics[width=1\columnwidth]{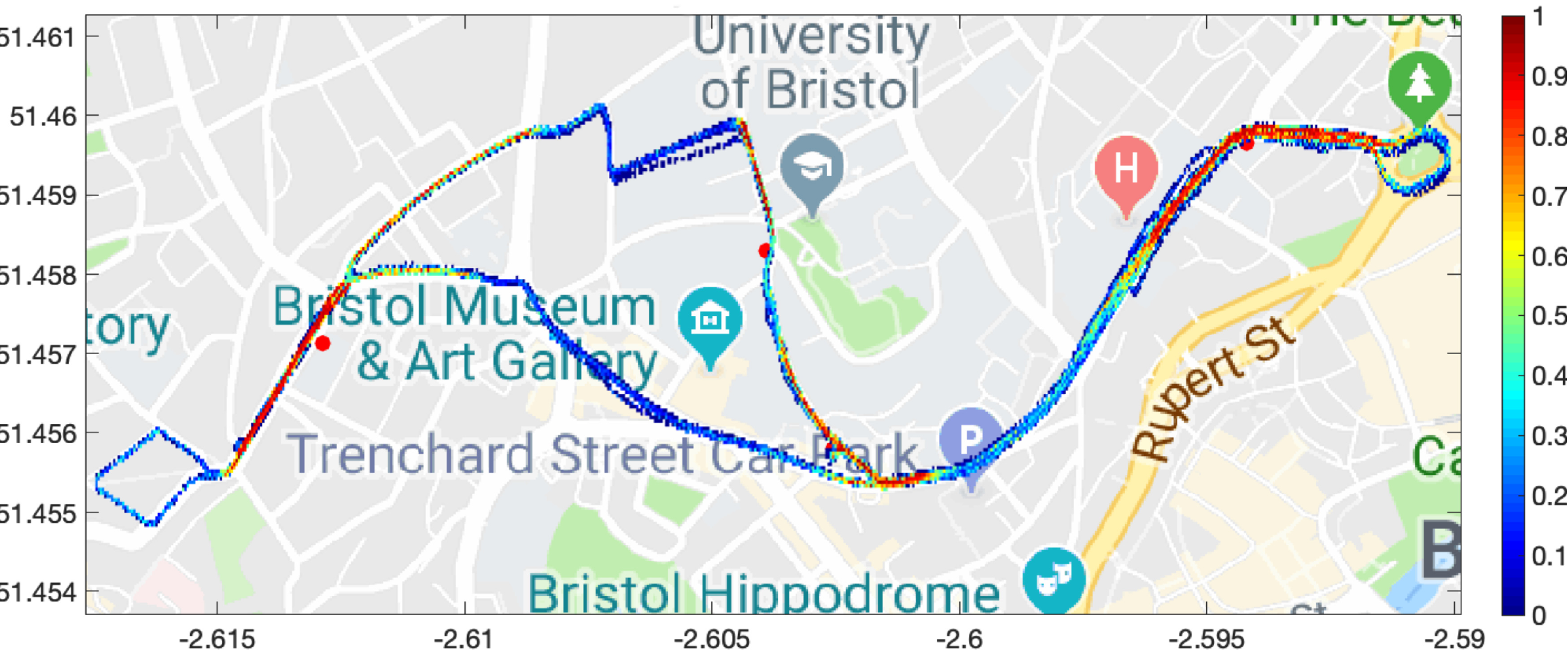}
    \caption{The heatmap results for all RSUs in our system. This figure shows the packet delivery rate for all RLNC-CAMs transmitted from a vehicle to a RSU and all the positions of the RSUs.}
    \label{fig:heatmap}
\endminipage\hfill
\minipage{0.49\textwidth}
\centering
\centering
    \includegraphics[width=1\columnwidth]{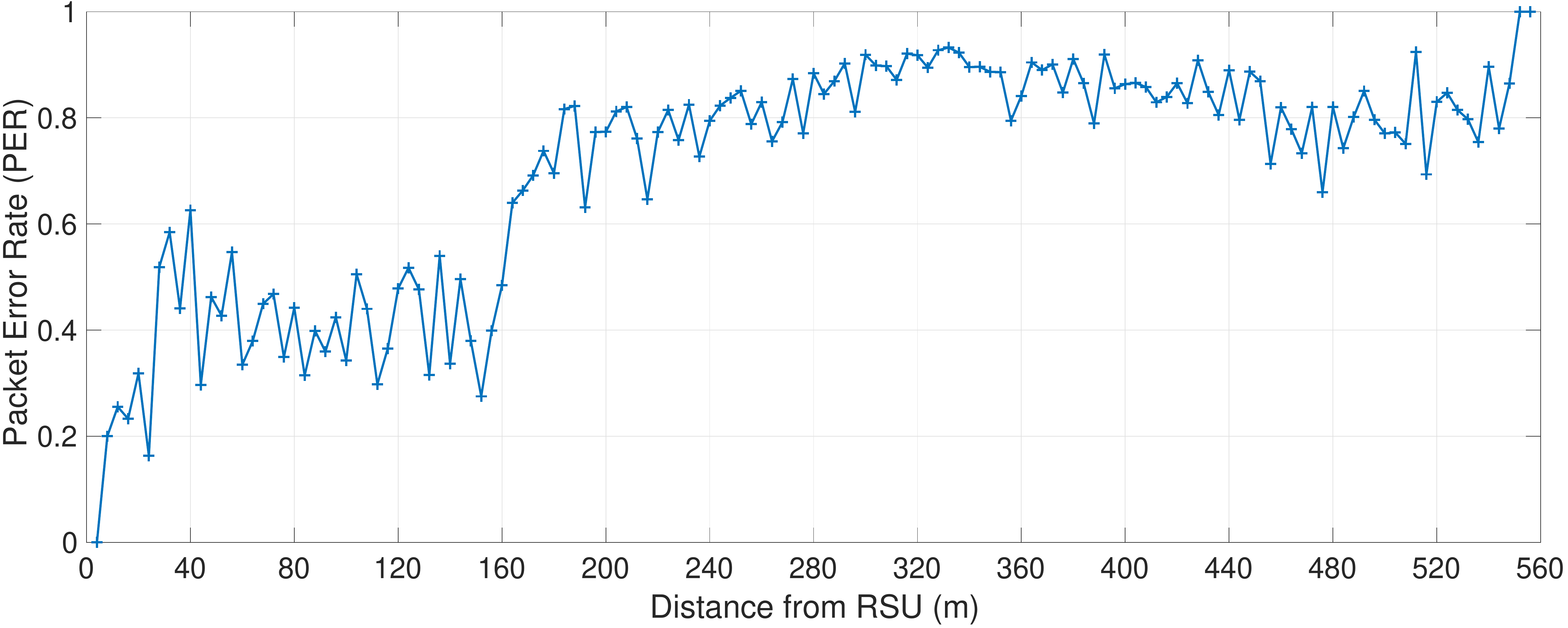}
    \caption{The packet error probability for \emph{RSU1} as a function of the distance. Each bar represents probability of a RLNC-CAM to be lost when a vehicle is within this particular $\SI{4}{\meter}^2$ area and the given distance from the RSU.}
    \label{fig:perPlot}
\endminipage\hfill
\end{figure*}

%The proposed RLNC-CAM differs from the standard CAM in 
The RLNC fields that follow the optional CAM attributes are defined as follows:
\begin{itemize}
    \item \emph{Source Message ID} contains the ID of the source message that is being transmitted. Similarly to the Station ID filed, this field is $32$ bits long, and it is defined as an integer value ranging between $0$ and $2^{32}-1$.
    \item \emph{Finite Field Size} encodes the value of $q$ that is being used by the RLNC-Facility Sublayer to generate each coded packet. In an embedded application, finite fields with a characteristic equal to $2$ are generally preferred because of the high level of optimization that it is possible to achieve in the implementation of the RLNC encoder/decoder~\cite{6227434}. Thus, for practical application, values of $q$ are restricted to $2, 2^2, 2^3, \ldots$. We propose that Finite Field Size field represents the value $\log_2(q) - 1$. Thus, if we limit the value of $q$ to $2^8$, it is sufficient for the field to be $3$ bits long to encode the values $2, 2^2, \ldots, 2^8$.
    \item \emph{Coding Seed}: Contains the seed used to initialize the pseudo-random number generator (PRNG) used to generate the coding vector associated with a coded packet. As such, provided the receiving end is equipped with the same PRNG, each coding vector can be precisely recovered. This solution is more efficient than including into each RLNC-CAM the entire coding vector represented as an array of integer values. Only a single integer representing the seed is included. By following the same reasoning as in~\cite{6353397}, we suggest the Coding Seed field to be $32$ bits long -- thus capable of expressing integer values ranging between $0$ and $2^{32}-1$.
    \item \emph{Coded Packet}: Contains the actual coded packet generated by the RLNC-Facility sublayer. Its bit length is variable and is specified by the RLNC-Facility sublayer.
\end{itemize}

As soon as an RSU receives an RLNC-CAM, it is forwarded to the FO and then to the cloud. Then, for any CAV and any source message with a given ID, a could-based service is responsible for checking if a number of RLNC-CAM carrying linearly independent coded packets have been received. If so, then an that particular source message can be recovered and forwarded to the other could-based services.

\section{Performance Evaluation}\label{sec:perfEval}
%The performance investigation for the above mentioned system is described in this section.
\begin{table}[t]
\renewcommand{\arraystretch}{1.07}
\centering
    \caption{Simulation and Experimental Parameters.}
    \begin{tabular}{r|rl}

    \textbf{Parameter}           & \textbf{Value}    & \\ \hline \hline
    Experiment/Simulation Time   & \SI{8}           & \SI{}{\hour}  \\
    Carrier Frequency            & \SI{5.9}          & \SI{}{\giga\hertz}  \\
    Bandwidth            & \SI{10}          & \SI{}{\mega\hertz}  \\
    RLNC-CAM Length            & \SI{2048}         & \SI{}{\byte}  \\
    RLNC-CAM  Interval              & \SI{10}          & \SI{}{\milli\second}  \\
    Transmission Power             &  \SI{29} & \SI{}{\dBm} \\
    Background Noise             & $\mathcal{N}\left( -110, 3 \right)$ & \SI{}{\dBm} \\
    Connector and Cable Losses   & 3                 & \SI{}{\dBm}
    \end{tabular}
\label{tab:parameters}
\end{table}

\begin{figure*}[t!]
\minipage{0.49\textwidth}
\centering
	\includegraphics[width=1\columnwidth]{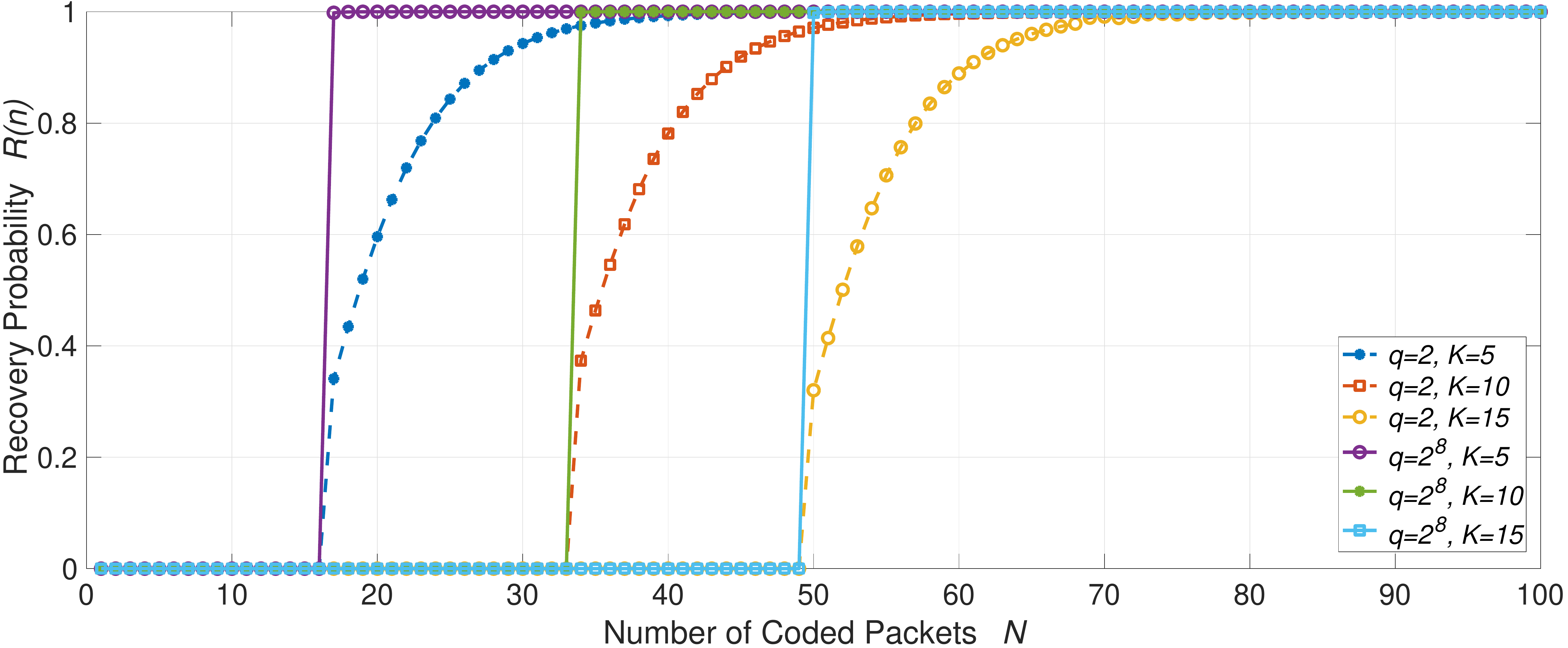}
    \caption{The recovery probability $R$ as a function of the number of RLNC-CAM transmissions, for $K = 5, 10, 15$ and $q = 2, 2^8$ as received by \emph{RSU1}.}
    \label{fig:rsu1RecProb}
\endminipage\hfill
\minipage{0.49\textwidth}
\centering
\centering
    \includegraphics[width=1\columnwidth]{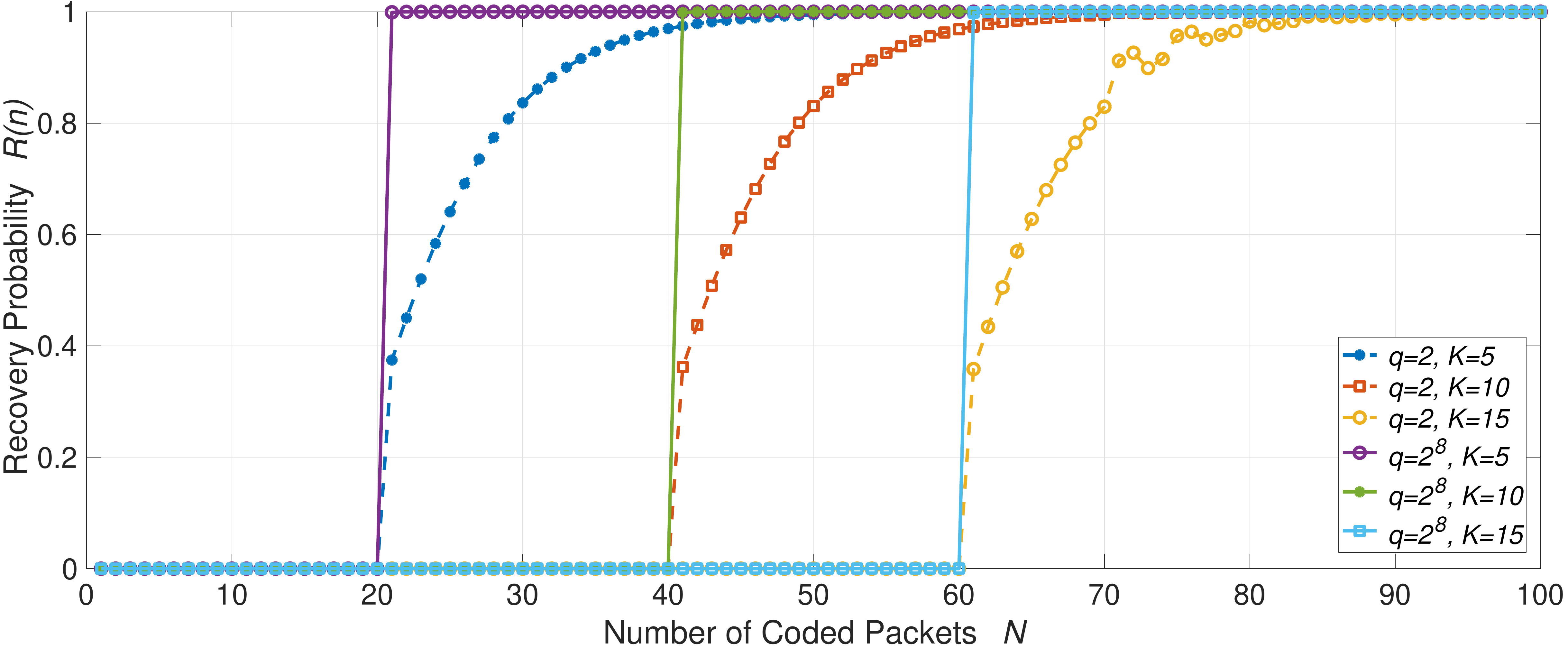}
    \caption{The recovery probability $R$ as a function of the number of RLNC-CAM transmissions, for $K = 5, 10, 15$ and $q = 2, 2^8$ as received by \emph{RSU2}.}
    \label{fig:rsu2RecProb}
\endminipage\hfill
\vspace{4mm}
\minipage{0.49\textwidth}
\centering
    \includegraphics[width=1\columnwidth]{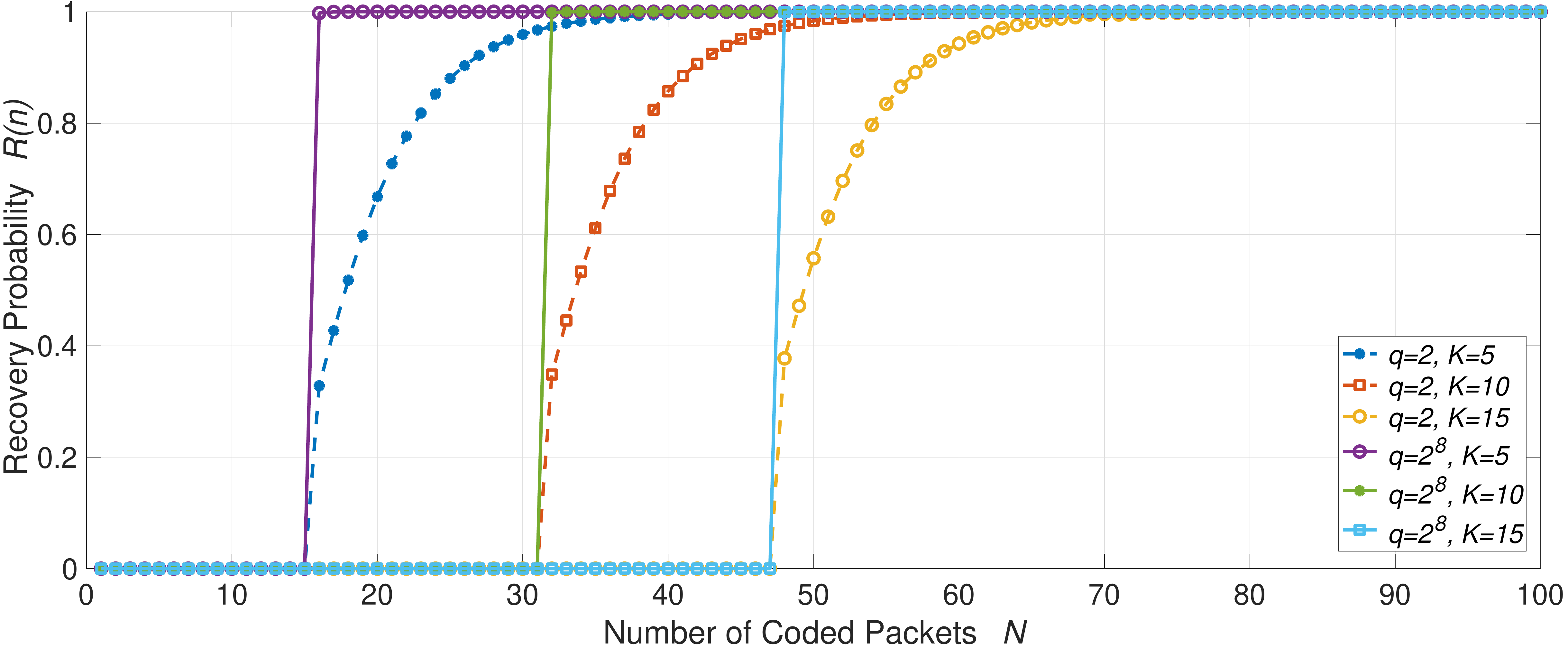}
    \caption{The recovery probability $R$ as a function of the number of RLNC-CAM transmissions, for $K = 5, 10, 15$ and $q = 2, 2^8$ as received by \emph{RSU3}.}
    \label{fig:rsu3RecProb}
\endminipage\hfill
\minipage{0.49\textwidth}
\centering
  \includegraphics[width=1\columnwidth]{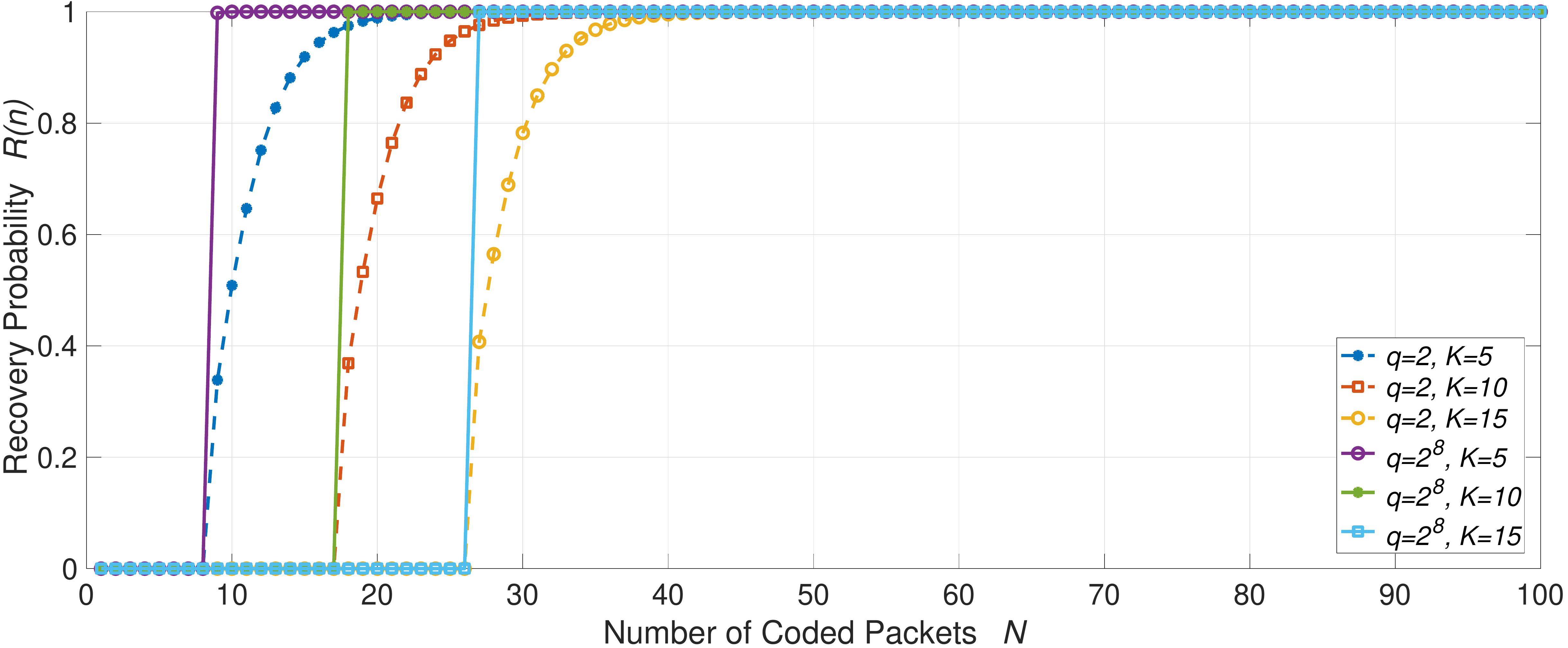}
    \caption{The recovery probability $R$ as a function of the number of RLNC-CAM transmissions, for $K = 5, 10, 15$ and $q = 2, 2^8$ as received by \emph{RSU4}.}
    \label{fig:rsu4RecProb}
\endminipage\hfill
\end{figure*}

We consider an infrastructure network composed of four RSUs, deployed in the City of Bristol, UK. More specifically, four RSUs were deployed (as shown in Fig.~\ref{fig:heatmap}). \emph{RSU1} was mounted close to a blind T-junction, on a curvy road at the height of \SI{8}{\meter}. The position of \emph{RSU2} was a straight road with light foliage, being mounted on the wall of a building at \SI{5}{\meter}. \emph{RSU3} was the highest placed RSU, mounted at the balcony at \SI{25}{\meter}, and providing coverage to a wide straight road. Finally, \emph{RSU4} can be located at one of the main arteries of the City of Bristol, being mounted at \SI{12}{\meter}. The different positions and buildings were chosen to investigate their effect in the performance of each individual link. Each RSU has a direct link to the Fog Computing Infrastructure described above. These deployed RSUs are being used for the two phases of our field-trials, with the first one being described in~\cite{adHocNowCityScale}.

During our experimental campaign, four vehicles were driven around the City of Bristol, UK for several hours, following the route shown in Fig.~\ref{fig:heatmap} (two in clockwise direction and two in anti-clockwise direction). All the transceivers operated at the frequency band of \SI{5.9}{\giga\hertz} and broadcast a RLNC-CAM per \SI{10}{\milli\second}. Each RLNC-CAM encapsulated the information shown in Fig.~\ref{fig:cam}. By means of a logging interface, we logged all the transmitted and received RLNC-CAMs. More information about the deployed testbed, the devices we utilized and the implemented communication stack, can be found in~\cite{adHocNowCityScale}. All the simulation and experimental parameters can be found in Table~\ref{tab:parameters}.
%The recovery probability curves for the RLNC integration were evaluated by means of Monte Carlo simulations. The exchanged packet error probability was derived from our real-world traces.

Starting with Fig.~\ref{fig:heatmap}, we present the heatmap for the packet delivery probability for all RLNC-CAMs transmitted from a vehicle and received by an RSU. Each square on this heatmap is $\SI{4}{\meter}^2$. We observe that, as expected, the packet delivery rate increases as we approach an RSU, it peaks when a vehicle passes by and fends off until being outside of coverage area. Similarly, Fig.~\ref{fig:perPlot} shows the RLNC-CAM Packet Error Rate (PER) for \emph{RSU1} as a function of the distance.
%Each mark on the plot represents the PER for each square f size $\SI{4}{\meter}^2$ that is found at the given distance from an RSU. 
%When more than one tiles are found at the given distance, the PER is averaged across all of them.
We see that when a CAV is in close proximity to an RSU $\left(\leq\SI{35}{\meter}\right)$, the PER is $\simeq$$20\%$. For distances between \SIrange{35}{160}{\meter}, the PER fluctuates around $\simeq$$40\%$. Finally, for greater distances, the PER values increase at the range of $75\% - 90\%$. As this figure is based on real-world data, we can observe that the PER does not increase smoothly, but it fluctuates. This is similar to what we observed in~\cite{adHocNowCityScale} and proves the necessity for a Fog Computing Infrastructure and the integration for RLNC principle in the ITS-G5 stack. The PER for the rest of the RSUs behaves similarly. More specifically, for \emph{RSU4} the performance is slightly better in terms of PER and compared to \emph{RSU1}, for \emph{RSU2} is slightly worse, while for \emph{RSU3} is almost identical.
The different plots for the rest of the RSUs will not be presented in this work, due to the limited space.

% rsu3 = MVB --> 1
% rsu2 = HW  --> 2
% rsu4 = SU  --> 3
% rsu1 = DH  --> 4

Figs.~\ref{fig:rsu1RecProb}--\ref{fig:rsu4RecProb} show the recovery probability $\mathrm{R}$ as a function of the number of transmitted coded packets. $\mathrm{R}$ was individually evaluated for the four RSUs deployed in order to investigate the effect the surrounding environment can introduce. Different source message lengths $K$ were evaluated, namely, $K \in \left\lbrace 5, 10, 15 \right\rbrace$, as well as different sets of $N$ coded packets, with $N \in \left\lbrace K, \ldots, 100 \right\rbrace$. Finally, two different lengths for the finite field coefficients $g_{i,j} $ where considered, being $q \in \left\lbrace 2, 2^8 \right\rbrace$. In order to validate our solution by testing multiple combinations of RLNC parameters, $\mathrm{R}$ was investigated by means of a Monte Carlo simulation, based on the real-world traces acquired from our field trials. In particular, by employing our experimental PER, we calibrated our full-stack network simulator as per~\cite{agileCalibration} and recreated the network scenario shown in Fig.~\ref{fig:heatmap}.

% Add a table with the sim parameters
% Mention about the full stack simulator - we feed the traces there and we run a simulation with one vehicle - same loop.
% What is called CAM --> RLNC - CAM

Fig.~\ref{fig:rsu1RecProb}, for both values of $q$, shows that, for $K = 5, 10$ and 15, at least $17,34$ and $50$ packets are required to have a chance of recovering a RLNC-CAM, respectively. The values for the rest of the RSUs are $21,41,61$ for \emph{RSU2}, $16,32,48$  for \emph{RSU3}, and finally, $9,18,27$ for \emph{RSU4}. The difference between the four RSUs arises from the different positions and environments where they are located. As we commented before, \emph{RSU4} achieved the best PER performance, being the RSU positioned on a wide road. As shown, this is reflected in the number of packets required for the recovery of a source message, as well.

In general, as expected, for $q=2$, we observe that the number of RLNC-CAM transmissions needed to recover a source message is larger than or equal to the case when $q=2^8$. As noted in~\cite{7335581}, larger values of $q$ increase the computational complexity of the RLNC decoder. This is generally a problem in mobile devices with limited computational resources. In our case, the aforementioned issue does not apply as a cloud-based service is responsible for performing the RLNC decoding operations -- thus, it is reasonable to assume to have high performance computing capabilities at our disposal. For these reasons, our results lead us to recommend a value of $q$ equal to $2^8$.

%A higher order finite field impairs the gradual recovery, and RSUs manage to recover a message with less transmitted packets. However, the implication of that is the increased complexity the recovery. The results show that the number of transmitted packets can be adjusted with respect to the channel conditions to achieve a balance between the probability of recovering a message, the probability of eavesdropper being unable to reconstruct the message and the complexity of the system. Depending on the computational capabilities of the nodes, the finite field can be increased to improve the probability of recovering the entire message.

% remove the thing that compares the different K's with the finite field. 
% the line that is not smooth ---> mention this happens because of the real-world traces

\section{Conclusions}\label{sec:conclusions}
This paper addressed the pressing concern of providing a city-scale Fog Computing architecture enabling the offloading of vehicular sensor data. We propose a Fog Computing infrastructure interconnecting a number of RSUs that in turn, collected sensor data transmitted by CAVs. Both RSUs and CAVs communicate by means of the ETSI's ITS-G5 communication stack that we extended by integrating an RLNC-Facility Sublayer into the ITS-G5 Facility layer. We also propose CAVs to offload their sensor data in a network coded fashion employing a novel type of CAM (namely, the RLNC-CAM). Building upon channel conditions measured during large-scale car trials and utilizing Monte Carlo simulations, we investigated both the feasibility and effectiveness of our proposal.

\section*{Acknowledgment}
This work is part of the FLOURISH Project, which is supported by Innovate UK, under Grant 102582.

\bibliographystyle{IEEEtran}
\bibliography{IEEEabrv,papers}

\begin{thebibliography}{10}
\providecommand{\url}[1]{#1}
\csname url@samestyle\endcsname
\providecommand{\newblock}{\relax}
\providecommand{\bibinfo}[2]{#2}
\providecommand{\BIBentrySTDinterwordspacing}{\spaceskip=0pt\relax}
\providecommand{\BIBentryALTinterwordstretchfactor}{4}
\providecommand{\BIBentryALTinterwordspacing}{\spaceskip=\fontdimen2\font plus
\BIBentryALTinterwordstretchfactor\fontdimen3\font minus
  \fontdimen4\font\relax}
\providecommand{\BIBforeignlanguage}[2]{{%
\expandafter\ifx\csname l@#1\endcsname\relax
\typeout{** WARNING: IEEEtran.bst: No hyphenation pattern has been}%
\typeout{** loaded for the language `#1'. Using the pattern for}%
\typeout{** the default language instead.}%
\else
\language=\csname l@#1\endcsname
\fi
#2}}
\providecommand{\BIBdecl}{\relax}
\BIBdecl

\bibitem{qosRequirements}
M.~Agiwal, A.~Roy, and N.~Saxena, ``{Next Generation 5G Wireless Networks: A
  Comprehensive Survey},'' \emph{IEEE Commun. Surveys Tuts.}, vol.~18, no.~3,
  pp. 1617--1655, Sep. 2016.

\bibitem{hetNet}
K.~Zheng, Q.~Zheng, H.~Yang, L.~Zhao, L.~Hou, and P.~Chatzimisios, ``{Reliable
  and Efficient Autonomous Driving: The Need for Heterogeneous Vehicular
  Networks},'' \emph{IEEE Commun. Mag.}, vol.~53, no.~12, pp. 72--79, Dec.
  2015.

\bibitem{TassimmWave}
A.~Tassi, M.~Egan, R.~J. Piechocki, and A.~Nix, ``Modeling and design of
  millimeter-wave networks for highway vehicular communication,'' \emph{IEEE
  Transactions on Vehicular Technology}, vol.~66, no.~12, pp. 10\,676--10\,691,
  Dec. 2017.

\bibitem{machineLearning}
M.~Levi, Y.~Allouche, and A.~Kontorovich, ``{Advanced Analytics for Connected
  Car Cybersecurity},'' in \emph{Proc. of IEEE VTC-Spring 2018}, June 2018, pp.
  1--7.

\bibitem{fogArchitecture}
M.~Sookhak, F.~R. Yu, Y.~He, H.~Talebian, N.~S. Safa, N.~Zhao, M.~K. Khan, and
  N.~Kumar, ``{Fog Vehicular Computing: Augmentation of Fog Computing Using
  Vehicular Cloud Computing},'' \emph{IEEE Veh. Technol. Mag.}, vol.~12, no.~3,
  pp. 55--64, Sep. 2017.

\bibitem{fogComputing}
C.~Huang, R.~Lu, and K.~K.~R. Choo, ``{Vehicular Fog Computing: Architecture,
  Use Case, and Security and Forensic Challenges},'' \emph{IEEE Commun. Mag.},
  vol.~55, no.~11, pp. 105--111, Nov. 2017.

\bibitem{anomalyDetectionActions}
M.~Riveiro, M.~Lebram, and M.~Elmer, ``{Anomaly Detection for Road Traffic: A
  Visual Analytics Framework},'' \emph{IEEE Trans. Intell. Transp. Syst.},
  vol.~18, no.~8, pp. 2260--2270, Aug 2017.

\bibitem{networkCoding}
H.~Niu, M.~Iwai, K.~Sezaki, L.~Sun, and Q.~Du, ``{Exploiting Fountain Codes for
  Secure Wireless Delivery},'' \emph{IEEE Commun. Lett.}, vol.~18, no.~5, pp.
  777--780, May 2014.

\bibitem{networkCodingVehicular}
G.~G. M.~N. Ali, M.~Noor-A-Rahim, P.~H.~J. Chong, and Y.~L. Guan, ``{Analysis
  and Improvement of Reliability Through Coding for Safety Message Broadcasting
  in Urban Vehicular Networks},'' \emph{IEEE Trans. Veh. Technol.}, vol.~67,
  no.~8, pp. 6774--6787, Aug 2018.

\bibitem{etsiStandard}
``{EN 302 663},'' ETSI, Tech. Rep., 2013.

\bibitem{6416071}
E.~Magli, M.~Wang, P.~Frossard, and A.~Markopoulou, ``{Network Coding Meets
  Multimedia: A Review},'' \emph{IEEE Trans. Multimedia}, vol.~15, no.~5, pp.
  1195--1212, Aug 2013.

\bibitem{7335581}
A.~Tassi, I.~Chatzigeorgiou, and D.~E. Lucani, ``{Analysis and Optimization of
  Sparse Random Linear Network Coding for Reliable Multicast Services},''
  \emph{IEEE Trans. Commun.}, vol.~64, no.~1, Jan. 2016.

\bibitem{6979970}
A.~Festag, ``{Cooperative Intelligent Transport Systems Standards in Europe},''
  \emph{IEEE Commun. Mag.}, vol.~52, no.~12, pp. 166--172, Dec. 2014.

\bibitem{etsiCam}
``{EN 302 637-2},'' ETSI, Tech. Rep., 2014.

\bibitem{6227434}
T.~Mladenov, S.~Nooshabadi, and K.~Kim, ``{Efficient GF(256) raptor code
  decoding for multimedia broadcast/multicast services and consumer
  terminals},'' \emph{IEEE Trans. Consum. Electron.}, vol.~58, no.~2, pp.
  356--363, May 2012.

\bibitem{6353397}
C.~Khirallah, D.~Vukobratovic, and J.~Thompson, ``{Performance Analysis and
  Energy Efficiency of Random Network Coding in LTE-Advanced},'' \emph{IEEE
  Trans. Wireless Commun.}, vol.~11, no.~12, pp. 4275--4285, Dec. 2012.

\bibitem{adHocNowCityScale}
I.~Mavromatis, A.~Tassi, R.~J. Piechocki, and A.~Nix, ``{A City-Scale ITS-G5
  Network for Next-Generation Intelligent Transportation Systems: Design
  Insights and Challenges},'' in \emph{Ad-hoc, Mobile, and Wireless Networks},
  N.~Montavont and G.~Z. Papadopoulos, Eds.\hskip 1em plus 0.5em minus
  0.4em\relax Cham: Springer International Publishing, 2018, pp. 53--63.

\bibitem{agileCalibration}
------, ``{Agile Calibration Process of Full-Stack Simulation Frameworks for
  V2X Communications},'' in \emph{Proc. of IEEE VNC 2017}, Nov 2017, pp.
  89--96.

\end{thebibliography}
\end{document}